\documentclass[12pt, prd, showpacs]{revtex4}
\input{tcilatex}

\begin{document}

\title{Traversable wormholes: minimum violation of null energy condition
revisited}
\author{O. B. Zaslavskii}
\affiliation{Astronomical Institute of Kharkov V.N. Karazin National University, 35
Sumskaya St., Kharkov, 61022, Ukraine}
\email{ozaslav@kharkov.ua}

\begin{abstract}
It was argued in literature that traversable wormholes can exist with
arbitrarily small violation of null energy conditions. We show that if the
amount of exotic material near the wormhole throat tends to zero, either
this leads to a horn instead of a wormhole or the throat approaches the
horizon in such a way that infnitely large stresses develop on the throat.
\end{abstract}

\pacs{04.70.Dy, 04.62.+v,}
\maketitle

% It is always \today, today, but any date may be explicitly specified


\section{Introduction and basic equations}

One of the key features of wormholes consists in that the null energy
condition (NEC) should be violated in the vicinity of a throat \cite{th}.
The corresponding material has the property unusual for laboratory physics
and is called "exotic". Although the inevitability of violation of NEC is
well known in physics of wormholes, the extent to which it occurs is still
being debated in literature. It was stated in \cite{sm} that amount of
exotic matter needed to support wormholes can be made as small as one likes
(see also \cite{kalam} - \cite{das}). Let we have the sequence of
traversable wormholes depending of some parameter $\varepsilon $ such that
in the limit $\varepsilon \rightarrow 0$ the amount of exotic material tends
to zero. According to \cite{sm}, one is inclined to think that a
configuration with arbitrary small but non-zero $\varepsilon $ represents an
usual traversable wormhole and, thus, some kind of discontinuity happens
that separates the state with $\varepsilon =0$ from those with $\varepsilon
\neq 0$. By itself, this does not necessarily mean something wrong since
topological properties of the configuration with $\varepsilon =0$ and $%
\varepsilon \neq 0$ are qualitatively different, so one may or may not
expect discontinuity. Indeed, a state with small $\varepsilon \neq 0$
connects two asymptotically flat universes whereas the state with $%
\varepsilon =0$ does not.

Nonetheless, the work \cite{sm} left one key question open. Let us call, for
brevity, a wormhole standard if 1) there is no horizon, 2) its geometry
remains regular (Kretschmann scalar is finite), 3) the areal radius $r$ as a
function of the proper distance $l$ grows away from the throat in some
finite vicinity. It is not quite clear from the results of \cite{sm} whether
or not the limiting configuration remains the standard wormhole when the
amount of exotic material tends to zero. The aim of the present article is
to show that the answer to this question is negative. In this sense, if we
restrict ourselves by standard wormholes, an amount of the exotic material 
\textit{cannot} be made arbitrarily small. We also find what metric appears
in the limit $\varepsilon =0$ depending on which of conditions 1) - 3) is
violated.

It is worth noting another approach in which the constraints on the wormhole
geometries are derived from quantum inequalities which null-contracted
stress-energy tensor should obey, when averaged over a timelike worldline 
\cite{fr}. Then, it turns out that in the concrete models considered as
candidates for arbitrarily small violation of NEC, the typical throat radius
cannot be macroscopic. Meanwhile, in our approach we do not constraint the
properties of the source (so, classical sources are, in principle, also
allowed provided they are compatible with NEC violation) and appeal directly
to geometrical consequences which follow directly from the definitions of
wormholes and Einstein equations.

We consider spherically-symmetric metrics%
\begin{equation}
ds^{2}=-e^{2\Phi }dt^{2}+\frac{dr^{2}}{V}+r^{2}d\omega ^{2}\text{, }d\omega
^{2}=\sin ^{2}\theta d\phi ^{2}+d\theta ^{2}\text{, }V=1-\frac{b}{r}
\label{m}
\end{equation}%
where $b$ and $\Phi $ are the shape and redshift functions, respectively. We
assume that the stress-energy tensor has the diagonal form%
\begin{equation}
T_{\mu }^{\nu }=diag(-\rho ,p_{r},p_{t},p_{t})\text{.}
\end{equation}

Then, it follows from the Einstein equations that%
\begin{equation}
\rho =\frac{b^{\prime }}{8\pi r^{2}}\text{; }p_{r}=\frac{1}{8\pi }[-\frac{b}{%
r^{3}}+2(1-\frac{b}{r})\frac{\Phi ^{\prime }}{r}]\text{; }p_{t}=\frac{1}{%
8\pi }\{(1-\frac{b}{r})[\Phi ^{\prime \prime }+\Phi ^{\prime }(\Phi ^{\prime
}+\frac{1}{r})]-\frac{b^{\prime }-\frac{b}{r}}{2r}(\Phi ^{\prime }+\frac{1}{r%
})\}\text{.}  \label{eq}
\end{equation}

The $r-r$ component of the conservation law $T_{\mu ;\nu }^{\nu }=0$ that
can be obtained from Einstein equations (\ref{eq}) reads%
\begin{equation}
p_{r}^{\prime }+\Phi ^{\prime }\rho +(\Phi ^{\prime }+\frac{2}{r})p_{r}-%
\frac{2p_{t}}{r}=0\text{.}  \label{cl}
\end{equation}

It is convenient to introduce the quantity

\begin{equation}
\xi \equiv 8\pi r^{2}(p_{r}+\rho )\text{ }  \label{ksi}
\end{equation}%
and use, along with the coordinate $r$, also the proper distance%
\begin{equation}
l=\int \frac{dr}{\sqrt{V}}\text{.}  \label{l}
\end{equation}

Then one obtains from (\ref{m}), (\ref{eq}) that%
\begin{equation}
V=1-\frac{r_{0}}{r}-\frac{2m(r)}{r}\text{, }m(r)=4\pi \int_{r_{0}}^{r}\rho 
\bar{r}^{2}d\bar{r}\text{,}  \label{v}
\end{equation}%
\begin{equation}
\xi =-V^{\prime }r+2\Phi ^{\prime }rV\text{, }  \label{zb}
\end{equation}%
\begin{equation}
\frac{d^{2}r}{dl^{2}}=\frac{V^{\prime }(r)}{2}\text{.}  \label{r2v}
\end{equation}

At the throat $r=r_{0}$ we must have, by definition, $\frac{dr}{dl}=0$, so
that $b(r_{0})=r_{0}$, and%
\begin{equation}
\frac{d^{2}r}{dl^{2}}(r_{0})=-\frac{\xi (r_{0})}{2r_{0}}\text{,}  \label{2r0}
\end{equation}%
\begin{equation}
p_{t}(r_{0})=\frac{1-b^{\prime }(r_{0})}{16\pi r_{0}}[\Phi ^{\prime }(r_{0})+%
\frac{1}{r_{0}}]\text{.}  \label{pt0}
\end{equation}

We are interested in traversable wormholes, so the horizon is supposed to be
absent, $\Phi (r_{0})$ is finite.

\section{Measuring amount of exotic matter}

Now we must choose the method to measure the degree of "exoticism". In
general, NEC is violated on the throat and/or in some vicinity of it.
Therefore, one can, instead of NEC itself, consider averaged null condition
(ANEC) obtained by integration of NEC \cite{book}. This procedure was
somewhat changed in \cite{sm}. To gain information about the total amount of
matter which violates NEC, it was suggested in \cite{sm} to consider the
volume integral $I\equiv \int_{r_{0}}^{\infty }dr\xi $. In this connection,
we must make a technical remark. The boundary term at infinity was lost in
eq. (12) of \cite{sm} which is equal to $b(\infty )$ and, in general, does
not vanish contrary to what is stated in \cite{sm}. What is more important,
the definition of $I$ should be, in our view, modified. In the form it was
introduced in \cite{sm}, the integral extends to infinity. Therefore, it may
happen that $I>0$ but there is a region of strong violation of NEC
compensated by the contribution of the normal matter. Meanwhile, it looks
more natural to be interested in the contribution from the exotic region
alone. Therefore, we will consider somewhat different definition%
\begin{equation}
I\equiv \int_{r_{0}}^{a}dr\xi =8\pi \int_{r_{0}}^{a}drr^{2}(p_{r}+\rho )%
\text{,}  \label{i}
\end{equation}%
where it is assumed that the exotic matter fills the inner region $r_{0}\leq
r<a$, while the outer region $r\geq a$ is occupied by the normal matter with
NEC satisfied or by vacuum ("normal region" for brevity). In doing so, we
use the same integral measure as in \cite{sm} but restrict the integration
by the exotic region. The appearance of the factor $drr^{2}$ in the integral
(\ref{i}) is motivated by the analogy with the ADM mass formula as is
explained in \cite{sm}. For our system, the quantity $\xi <0$ for $r_{0}\leq
r<a$ and $\xi \geq 0$ for $r\geq a$. Actually, in concrete examples
considered in \cite{sm} the value of $I$ does not depend on which of two
definitions is used since the outer region lies in vacuum and does not
contribute to the integral $I$.

As $\xi \leq 0$ in the integrand, we have $I\leq 0$. Our goal is to
elucidate whether and under which conditions \ one can obtain $I\rightarrow
0 $. As we want to preserve regularity we consider everywhere finite $\xi $.
Then $I\rightarrow 0$ entails that either 1) $a\rightarrow r_{0}$ (limits of
integration shrink) or 2) $\xi \rightarrow 0$ (the integrand vanishes) in
the whole exotic region (or both 1 and 2 hold). In other words, we can try
to minimize either the size of the region or "exoticism" or violation of NEC
in the relationship between $p_{r}$ and $\rho $ (i.e. in the equation of
state). Let us discuss different cases separately.

\section{Minimizing size of region with exotic material}

\subsection{Continuous distribution}

\subsubsection{$\protect\xi (r_{0})<0$}

Let us now assume that pressures $p_{r}$, $p_{t}$ and the energy density $%
\rho $ are continuous functions of $r$. Therefore, on the border between the
exotic and normal regions,%
\begin{equation}
\xi (a)=0\text{.}  \label{x}
\end{equation}

On the throat, it is supposed that NEC are violated as usual \cite{th}, \cite%
{book}, so $\xi (r_{0})<0$. We are interested in regular configurations. The
function $\Phi (r)$ and its first and second derivatives are supposed to be
bounded, unless the opposite is stated explicitly. In the limit under
discussion lim$_{a\rightarrow r_{0}}\frac{d\xi }{dr}(a)=\lim_{a\rightarrow
r_{0}}\frac{\xi (a)-\xi (r_{0})}{a-r_{0}}\rightarrow \infty $. It follows
also from (\ref{cl}) that $p_{r}^{\prime }$ is finite. Then, the only way to
reconcile the finiteness of $p_{r}^{\prime }$ with divergency of $\xi
^{\prime }$ consists in considering infinite $\rho ^{\prime }$. Let us
consider, without a big loss of generality, the density profile of the form%
\begin{equation}
\rho =\rho _{0}+\Delta \rho \frac{(r-r_{0})^{n}}{(a-r_{0})^{n}}  \label{ror}
\end{equation}%
for $r_{0}\leq r\leq a$ and $\rho (r)$ is some smooth function with bound
derivatives for $r>a$. It follows from (\ref{ror}) that $\rho (r_{0})=\rho
_{0}$, $\rho (a)=\rho _{0}+\Delta \rho $. From (\ref{x}), we have that $\rho
(a)=-p_{r}(a)$. In the limit $a\rightarrow r_{0}$ $\rho ^{\prime
}(a)\rightarrow \infty $, so we have a jump in $\rho $. (For $a-r_{0}$ small
but non-zero the distribution of $\rho $ is still continuous.) Due to
continuity of $p_{r}$ and the finiteness of $p_{r}^{\prime }$, in this limit
we obtain also that $\rho (a)=-p_{r}(r_{0})+O(\delta )$, $\delta =a-r_{0}$.
On the throat, eq. (\ref{eq}) entails a well-known equality $%
p_{r}(r_{0})=-(8\pi r_{0}^{2})^{-1}$ \cite{th}, \cite{book}. Therefore, for
small $\delta $%
\begin{equation}
\rho (a)=\frac{1}{8\pi a^{2}}+O(\delta )\text{.}  \label{rd}
\end{equation}

How does the geometry look like in this limit? To answer this question,
consider first the geometry for $r>a$. We assume that for $r\rightarrow a$
the corresponding function admits the Taylor expansion $V=V(a)+V^{\prime
}(a)(r-a)+\frac{V^{\prime \prime }(a)}{2}(r-a)^{2}+O((r-a)^{3})$ with finite
coefficients. Then, it follows from (\ref{v}) and (\ref{rd}) that $%
V(a)\rightarrow 0$ and $V^{\prime }(a)\rightarrow 0$ in the limit $\delta
\rightarrow 0$. As a result, the metric in the vicinity of $r=a$ looks like%
\begin{equation}
ds^{2}=-dt^{2}\exp [2\Phi (a)]+\frac{2dr^{2}}{V^{\prime \prime }(a)(r-a)^{2}}%
+a^{2}d\omega ^{2}
\end{equation}%
and represent a horn in the sense that the proper distance between $r=a$ and 
$r>a$ diverges. In doing so, the quantity $\exp [2\Phi (a)]$ does not vanish
since there is no horizon by our assumption. The part of the manifold $r\geq
a$ becomes geodesically complete and does not resemble a wormhole.

It is also instructive to trace what happens to the region $r_{0}\leq r<a$
in this limit. Again, we assume the Taylor expandability in the vicinity of $%
r_{0}$. Then, it is easy to obtain that for small $\delta $ the expansion
reads%
\begin{equation}
V=\frac{r-r_{0}}{r}Z\text{,}  \label{vz}
\end{equation}%
\begin{equation}
Z=-\xi (r_{0})-8\pi r_{0}^{2}\Delta \rho \frac{(r-r_{0})^{n}}{(n+1)\delta
^{n}}+O(\delta ^{2})\text{.}  \label{z}
\end{equation}

Making a substitution $r=r_{0}+\delta y$, where $0\leq y\leq 1$ we obtain
that%
\begin{equation}
ds^{2}=-dt^{2}\exp [2\Phi (a)]+\delta \frac{dy^{2}}{f(y)}+r_{0}^{2}d\omega
^{2}\text{,}
\end{equation}%
$f(y)=-\xi (r_{0})y-8\pi r_{0}^{2}\Delta \rho \frac{y^{n}}{(n+1)}$ does not
contain $\delta $. Then, it becomes obvious that the proper distance between
the throat and the boundary \thinspace \thinspace $r=a$ between the normal
and exotic regions is finite and, moreover, tends to zero, so that this part
of space shrinks to a disc and, thus, is removed from the manifold.

\subsubsection{Case $\protect\xi (r_{0})=0$}

Up to now, we assumed that $\xi (r_{0})<0$. Correspondingly, the asymptotic
expansion of the metric coefficient began from the linear terms according to
(\ref{z}). The situation changes, if $\xi (r_{0})=0$. The exotic region $\xi
(r)<0$ is assumed to occupy the interval \thinspace $r_{0}<r<a$ and $\xi
(a)=0$. Let we try again to find the configuration with the almost zero
violation of NEC. As now $\xi (r_{0})=0$, it follows from (\ref{zb}) that $%
V^{\prime }(r_{0})=0$. It follows from (\ref{v}) that this condition means $%
\rho =\rho _{0}=(8\pi r_{0}^{2})^{-1}$. We assume that in the vicinity of
the throat we have the asymptotics%
\begin{equation}
V=A(r-r_{0})^{n+1}
\end{equation}%
with $A=const$ and $n>0$. Instead of an usual minimum of the function $r(l)$
typical of $n=0$, now we deal with the minimum of the higher order. We want
to find a regular wormhole configuration so that we require that the proper
distance to the throat be finite. Thus, $0<n<1$. This is similar to the case
discussed in section 4.1 of \cite{das}. 

Then, we obtain from (\ref{zb}) that 
\begin{equation}
\xi =-r_{0}(n+1)A(r-r_{0})^{n}+B(r-r_{0})^{n+1}\text{\thinspace }+...
\end{equation}%
The exact value of the coefficient $B$ (as well as the coefficients at the
higher degrees in the expansion) is irrelevant for us. What is important is
the fact that $A$ does not contain the small parameter, so that $\xi $ has
the general form $\xi =-A_{1}y+A_{2}y^{m}$, $m=\frac{n+1}{n}>1$, $y=r-r_{0}$%
, $A_{1}$ and $A_{2}$ are finite, $A_{1}>0$. Then, it becomes clear that the
function $\xi $ attains its minimum at some $r_{1}$ with finite $r_{1}-r_{0}$%
. In turn, this means that $a-r_{0}>a-r_{1}$ is also finite and cannot be
made arbitrarily small. Thus, although now the proper distance is finite,
our efforts to achieve the almost zero violation of NEC failed again.

From the other hand, we may impose by brute force the condition $%
a\rightarrow r_{0}$, allowing sequence of configurations with more and more
small $A$. As the point $r=a$ is supposed to be a regular point, we may
exploit the Taylor expansion $V=V(a)+V^{\prime }(a)(r-a)+V^{\prime \prime
}(a)\frac{(r-a)^{2}}{2}+...$. It follows from the condition $\xi (a)=0$ and
eq. (\ref{zb}) that $V^{\prime }(a)\sim V(a)\rightarrow V(r_{0})=0$. Thus,
only the term proportional to $(r-a)^{2}$ may survive here, so the proper
distance to $r=a$ diverges.

The analysis becomes especially simple if $\Phi =0$. Then, we have from (\ref%
{zb}) that%
\begin{equation}
\xi =-V^{\prime }r\text{.}
\end{equation}%
Then, $V^{\prime }$ changes the sign at $r=a$, so that $V^{\prime }<0$
everywhere in a normal region. It means that $V(r)$ decreases to zero that
does not correspond to a wormhole configuration (by assumption, $V=0$ at the
throat $r_{0}$ only but $V>0$ for $r>r_{0}$).

In the work \cite{kuh2} the trial form of the shape function was chosen so
that $V=\exp [\frac{K}{(r-r_{0})^{n}}]$ with $n\geq 1$. Then, it is obvious
that $V$ diverges so strongly near the throat that the proper distance $%
l\rightarrow \infty $ and we have a horn instead of a wormhole.

\subsection{Jump of $\protect\xi $}

The above arguments do not work directly if we allow the jump of the
quantity $\xi (r)$. Let this quantity change from $\xi (a-0)<0$ in the
exotic region to $\xi (a+0)>0$ in the normal one. Then, \ for $a$ close to $%
r_{0}$ it follows from (\ref{zb}) that $V^{\prime }$changes by jump from
positive to negative values at $r=a$, while the term proportional to $V$ is
negligible. However, if the function $V(r)$ is small and negative near $%
a\rightarrow r_{0}$ and has a negative finite derivative $V^{\prime }=-\frac{%
\xi }{r}+O(V)$ for $r>a$ it means that the function $V$ changes the sign at
some $b=a+O(V)$ in obvious contradiction with the properties of wormhole
metrics. If, instead, $\xi $ changes by jump from $\xi (a-0)<0$ to $\xi
(a+0)=0$, the derivative $V^{\prime }(a)\rightarrow 0$ and we have a horn.

\section{Minimizing "exoticism"}

Let $\xi (r)\rightarrow 0$ inside some region including $r=r_{0}$. If $\xi
=0 $ (or, equivalently, $p_{r}=-\rho $) it follows from the Einstein
equations (\ref{eq}) that only two possibilities exist.

\subsection{Horizons instead of traversable wormholes}

First case: $\exp (2\Phi )=1-\frac{b}{r}$. In this case at the supposed
throat $r=r_{0}$ where $b(r_{0})=r_{0}$ the $g_{00}$ component of the metric
tensor in (\ref{m}) vanishes as well in contradiction with the assumption
about the absence of the horizon. Therefore, this case should be rejected.

\subsection{Horn-like configurations}

Second case: $b=r$. Then, formally, $g_{11}=\infty $. Actually, this only
means that the coordinate $r$ is degenerate and cannot be used. If, instead,
we use $l$, one can see that $\frac{dr}{dl}=0$, so $r=r_{0}=const$ inside
the corresponding region. Thus, instead of the wormhole we have a horn. In
principle, one can take a finite piece of this horn and glue it to the
metric with $\frac{dr}{dl}>0$ on the right side and $\frac{dr}{dl}<0$ on the
left side. In doing so, one can obtain so-called "null wormholes"
(N-wormholes) \cite{n}. Although inside the region where $p_{r}+\rho =0$ NEC
is satisfied on the verge, gluing to the normal matter on the borders needs
the presence of exotic material and we return to the situation under
discussion with all corresponding difficulties.

One can also consider the situation when $\xi (r)\rightarrow 0$ when $%
r\rightarrow r_{0}$ but does not vanish identically in some region. Then, $r$
is also not identically constant. In the vicinity of $r_{0}$ we still have a
horn (semi-infinite throat) with small derivatives of $r$. Indeed, it
follows directly from (\ref{zb}), that in this limit $\frac{d^{2}r}{dl^{2}}%
(r_{0})\rightarrow 0$. In a similar way, one can show, assuming analyticity
of $b(r)$, that all higher derivatives $\frac{d^{n}r}{dl^{n}}\rightarrow 0$.
Actually, it means that $r_{0}$ is at infinite proper distance from any $%
r>r_{0}.$ As a result, the spacetime with $r$ varying from $r_{0}$ to the
right infinity is geodiesically complete. Therefore, our configuration
cannot be considered as a wormhole.

As an example of such a configuration, we can point to the exact solutions
in dilaton gravity \cite{dil}:%
\begin{equation}
ds^{2}=-dt^{2}+(1-\frac{r_{0}}{r})^{-2}dr^{2}+r^{2}d\omega ^{2}\text{.}
\end{equation}%
In this example $\xi =-2\frac{r_{0}}{r}(1-\frac{r_{0}}{r})\rightarrow 0$
when $r\rightarrow r_{0}$, $r-r_{0}\simeq \exp (-\frac{l}{r_{0}})$ in this
limit. Meanwhile, $r$ is not identically constant and, moreover, $%
r\rightarrow \infty $ at asymptotically flat infinity. The spacetime with $%
r_{0}\leq r<\infty $ is geodesically complete and no wormhole arises.

One can also consider the case when $\xi (r_{0})\neq 0$ but is small. When,
for non-zero $\xi <0$ a wormhole configuration will be indeed possible.
However, in the limit $\xi (r_{0})\rightarrow 0$ it will be approaching the
horn-like one closer and closer, with $\frac{d^{2}r}{dl^{2}}(r_{0})$
becoming smaller and smaller, so in the limit $\xi (r_{0})=0$ we return to
the situation described above: either it will be a horn everywhere ($r=const$%
) or asymptotically for $r\rightarrow r_{0}$. Anyway, the limit of sequence
of such configuration does not represent a wormhole.

\subsection{Combined case}

One can also try to combine both factors and consider decreasing $\xi
(r_{0}) $ along with decreasing $a-r_{0}$. Actually, the model of this kind
was considered in \cite{kuh} where it is claimed that the violation of NEC
can be arbitrarily small but it was not discussed what happens to the
wormhole metric in this limit. In our notations, this model can be described
as follows. Let $\Phi =0$, $b=1-k-\epsilon (r)$ for $r_{0}\leq r\leq a$
where \thinspace $\epsilon ^{\prime }(r_{0})=0$, $k\rightarrow 1$, $%
a\rightarrow r_{0}$ and $\epsilon (r)$ has the order $k-1$ in this limit.
Then, one can obtains from \cite{kuh} or directly from (\ref{eq}) that $\rho
+p_{r}=-\frac{1-k}{8\pi r_{0}^{2}}<0$ inside the interval $[r_{0},a]$ where
it can be made arbitrarily small by taking sufficiently small $1-k$.
However, in this limit the proper distance between any $r>a$ and $r=a$
behaves like $l(a,r)=\frac{r-a}{\sqrt{1-k}}$ and diverges in this limit. As
a result, we again obtain a geodesically complete spacetime with a horn
instead of a wormhole. It was already mentoned in \cite{fr} that in this
model the proper distance between the throat and the border at $a$ behaves
similarly like $l(r_{0},a)=\frac{r_{0}-a}{\sqrt{1-k}}$, so that only
fine-tuning between parameters may warrant the finite $l(r_{0},a)$ in the
limit under discussion. We would like to point out that even such a
fine-tuning does not save the matter since $l(a,r)$ diverge even if $%
l(r_{0},a)$ remains finite. Shortcomings of this model as well as of some
other models also flaring outward too slowly (see \cite{fr} \ for their
detailed criticism) were repaired in \cite{k06} where minimizing the size of
exotic region was discussed without requirement of making this region
arbitrarily small. Instead, some balance between this size and restrictions
due to quantum inequalities \cite{fr},\cite{ford} was suggested. This issue,
however, is beyond the scope of the present paper.

\section{Limiting configurations with a horizon}

Our goal is to try to find wormhole configurations with the minimum
violation of NEC. We saw in previous sections that, typically, it is the
flare out condition which is violated in the limit under consideration, so a
wormhole becomes more and more extended and approaches the horn. It is seen
from (\ref{zb}) that the first term is negative near the throat while the
second one is positive and small, provided $\Phi ^{\prime }>0$ is finite.
Roughly speaking, what we did in previous section is the attempt to diminish
the first negative term by diminishing $V^{\prime }$. This attempt resulted
in the appearance of horns instead of throats since small $V^{\prime }$
entail large proper distances to the border between regions. Let us try
another method to achieve positive $\xi $ in the small vicinity of the
throat: instead of diminishing the first negative term we can try to
increase the second positive one. We assume that $V\sim r-r_{0}$ near the
throat. If we still take $\Phi "$ bounded near the throat the factor $V$
will make the second term in (\ref{zb}) negligible. Instead, we should take $%
\Phi \sim \ln (r-r_{0})$. As we want to have the throat, we need $\Phi $ to
be finite on the throat. The natural choice is%
\begin{equation}
\exp (\Phi )=\varepsilon +\lambda f(r)\text{, }r_{0}\leq r\leq a\text{, }%
\lambda >0  \label{bw}
\end{equation}%
which is glued smoothly to the patch with $r>a$. If $\varepsilon =0$ from
the very beginning, we obtain the horizon. Then, the requirement of
regularity along with the asymptotics $V\sim r-r_{0}$ leads to the condition 
$f\sim Af_{0}$, $f_{0}=\sqrt{1-\frac{r_{0}}{r}}\,\ $(by rescaling $\lambda $
we always may achieve $A=1$). For simplicity and without big loss of
generality, we restrict ourselves by the case when $f=f_{0}$, $V=1-\frac{%
r_{0}}{r}$ exactly and $\exp (\Phi )=\sqrt{V}$ for $r>a$. This is just the
example "Specialization 2" from \cite{sm}. The properties of this model were
also discussed in \cite{fr} where it was shown that quantum inequalities 
\cite{fr} constrain it in such a way that make it unrealistic. However, this
does not exclude, in principle, exploiting some unusual classical source for
this geometry. In our context, we are interested in inner properties of the
system irrespective of how it is created and do not appeal to any numerical
estimates.

In \cite{sm}, the authors mainly discussed the behavior of bulk density and
pressure, meanwhile now we will see that the crucial role is played by
surface stresses. They appear on the boundary between two different regions
"+" (right) and "-" (left) \cite{isr} with components $S_{0}^{0}$ and $%
S_{2}^{2}=S_{3}^{3}$ where%
\begin{equation}
8\pi S_{0}^{0}=\frac{2}{r}[\left( \frac{dr}{dl}\right) _{+}-\left( \frac{dr}{%
dl}\right) _{-}]\text{,}
\end{equation}%
\begin{equation}
8\pi S_{2}^{2}=\frac{1}{r}[\left( \frac{dr}{dl}\right) _{+}-\left( \frac{dr}{%
dl}\right) _{-}]+\left( \frac{d\Phi }{dl}\right) _{+}-\left( \frac{d\Phi }{dl%
}\right) _{-}\text{.}  \label{22}
\end{equation}

As $V$ is supposed to be continuous, $S_{0}^{0}=0$ everywhere. On the border 
$r=a$ 
\begin{equation}
8\pi S_{2}^{2}(a)=\frac{r_{0}\varepsilon }{2\varepsilon _{s}^{2}a^{2}}\text{%
, }\varepsilon _{s}\equiv \sqrt{1-\frac{r_{0}}{a}}  \label{sa}
\end{equation}%
where we follow the notations of \cite{sm}. We emphasize that for the metric
under discussion, there are also the surface stresses on the throat itself
where two branches of $r(l)$ with opposite signs of $\frac{dr}{dl}$ meet.
Simple calculations give us 
\begin{equation}
8\pi S_{2}^{2}(r_{0})=\frac{\lambda }{r_{0}\varepsilon }  \label{s0}
\end{equation}

Now we will show that the stresses (\ref{sa}) and (\ref{s0}) cannot be made
finite simultaneously in the limit $a\rightarrow r_{0}$. Indeed, in this
limit the quantity $\varepsilon _{s}\rightarrow 0$. If $\varepsilon $ is
fixed, $S_{2}^{2}(a)\rightarrow \infty $ that hardly can be accepted
physically. One may try to repair this shortcoming and, instead, take
simultaneously the limits $\varepsilon _{s}\rightarrow 0$, $\varepsilon
\rightarrow 0$ in such a way that $\frac{\varepsilon }{\varepsilon _{s}^{2}}$
remains finite or even zero. However, in the limit $\varepsilon \rightarrow
0 $, $S_{2}^{2}(r_{0})\rightarrow \infty $ independently of the relationship
between $\varepsilon $ and $\varepsilon _{s}$.

\section{Summary}

It turned out that the statement of the kind "there are wormholes with
arbitrarily weak violation of NEC" should be taken with great care. We have
shown that the construction of traversable wormholes with arbitrarily small
amount of the exotic material is achieved by expense of violation of some
properties inherent to standard traversable wormholes. We considered several
ways to achieve the arbitrarily small amount of exotic material for
wormholes geometry which, actually, can be divided to two types. The first
one consists in diminishing $V^{\prime }$(then the density approaches the
critical value $\rho _{0}=(8\pi r_{0}^{2})^{-1}).$ Then, typically, instead
of a wormhole we obtain \ a horn in this limit. It does not mean that such
configurations have no physical sense. By contrary, in some situations it is
the tube-like configurations which are placed to the forefront (for example,
in the context of the problem of avoidance singularities \cite{n}, in
investigating properties of phantom matter \cite{ph}, etc.). However, in any
case, they represent a special class of wormhole configurations which differ
from the standard ones. The second type represents wormholes on the
threshold of formation of a horizon. Then, infinitely large surfaces
stresses appear on the throat and/or the boundary between exotic and normal
regions, so the limit is singular and, in this sense, unphysical.

As the quantitative measure of "exoticism" we exploit the value of $I$
according to the definition (\ref{i}) used in \cite{sm}. If, instead of $I$
we would consider the similar integral $J$ over the proper distance $l$ we
would obtain that $\left[ J\right] >\left[ I\right] $ since $V<1$.
Therefore, as $\left[ I\right] $ cannot be made arbitrarily small without
violation of conditions 1) - 3) indicated in Introduction, the same applies
to $J$. In this sense, using $J$ as the quantifier of exoticism only
strengthens the conclusion that degree of exoticism cannot be made
arbitrarily small for standard wormholes.

One can think that if an advanced civilization wants to construct a standard
wormhole, it should not make lame excuses about "small amount" of exotic
material and must do inevitable job of collecting and arranging this
material in sufficient supply.

I am grateful to Sergey Sushkov for reading the manuscript and useful
comments.

\end{document}